\documentclass[letterpaper, 10 pt, conference]{ieeeconf}  

\IEEEoverridecommandlockouts                              

\usepackage{geometry}
\usepackage{amsthm}      
\usepackage{amsmath,amsfonts}
\usepackage{amssymb}  
\usepackage{algorithmic}
\usepackage{algorithm}
\usepackage{array} 
\usepackage{textcomp}
\usepackage{stfloats}
\usepackage{url}
\usepackage{verbatim}
\usepackage{graphicx}

\usepackage{subfloat}

\usepackage{subcaption}
\usepackage[font=small,labelfont=bf]{caption}
\usepackage[utf8]{inputenc} 

\usepackage{tabularx}

\usepackage{siunitx}
\DeclareMathOperator*{\argmin}{argmin} 

\usepackage{bondgraphs}
\usepackage{tikz}
\usetikzlibrary{calc}
\usepackage{epstopdf}
\usetikzlibrary{shapes,fit}

\newtheorem*{Note}{Note}
\newtheorem{theorem}{Theorem}
\newtheorem{lemma}{Lemma}
\newtheorem*{objective*}{PBC objective}
\newtheorem*{tankobjective*}{Sampled PBC objective}
\newtheorem{remark}{Remark}
\newtheorem{corollary}{Corollary}

\usepackage[style=ieee,
			url=false, 
			isbn=false,
			doi=true,
			giveninits=true,
			uniquename=init,
			isbn=false,
			backend=bibtex,
			minbibnames=1,
			maxbibnames=6	]{biblatex}

\bibliography{references.bib}

\title{\LARGE \bf
Passivity-Preserving Safety-Critical Control using Control Barrier Functions
}

\author{Federico Califano$^{1}$
\thanks{$^{1}$Robotics \& Mechatronics (RaM) group, University of Twente, The Netherlands.
        {\tt\small f.califano@utwente.nl}}
}

\begin{document}

\maketitle
\thispagestyle{empty}
\pagestyle{empty}

\begin{abstract}
In this letter we propose a holistic analysis merging the techniques of passivity-based control (PBC) and control barrier functions (CBF). We constructively find conditions under which passivity of the closed-loop system is preserved under CBF-based safety-critical control. The results provide an energetic interpretation of safety-critical control schemes, and induce novel passive designs which are less conservative than standard methods based on damping injection. The results are specialised to port-Hamiltonian systems and simulations are performed on a cart-pole system.

\end{abstract}

\section{Introduction}

\textit{Passivity-based control} (PBC) encompasses several techniques aiming to stabilise systems independently on external environmental interactions \cite{Ortega2001PuttingControl,vanderSchaftL2,Ortega2004InterconnectionSurvey,Stramigioli2015Energy-AwareRobotics}. These schemes use Lyapunov-like arguments to design closed-loop generalised energy functions (or storage functions) encoding both desired behaviors and stability guarantees for the controlled system \cite{Ortega2002InterconnectionSystems}.
A seemingly unrelated control tool is represented by \textit{safety-critical control}, a technique producing forward invariance of a \textit{safe set}, a subset of the state space defined as the superlevel set of so-called \textit{control barrier functions} (CBFs) \cite{Ames2019ControlApplications,Ames2017ControlSystems,Xu2015RobustnessControl}. Safety-critical control is practically implemented via solving a quadratic program minimising the distance from a desired control input, and as such producing a \textit{filtered} version of the control input which guarantees forward invariance of the safe set.

In this letter we investigate under which conditions this safety-critical filtering algorithm preserves passivity of the underlying controlled system, assuming that the desired input comes from a PBC design.
We specialise the results to the class port-Hamiltonian (pH) systems \cite{Duindam2009ModelingSystems}, encompassing for a great variety of physical systems including the totality of the mechanical ones. Due to its explicit display of energetic information, this formulation is very convenient when PBC schemes are developed \cite{Ortega2001PuttingControl,Ortega2002InterconnectionSystems}. It will be shown how the pH formulation used in a safety-critical framework induces intuitive and technical advantages with respect to a Lagrangian formulation, normally used in this context. 
As a consequence safety-critical control schemes gain a clear energetic interpretation, which can be used for multiple purposes in energy-aware schemes \cite{Stramigioli2015Energy-AwareRobotics,Ortega2001PuttingControl}. In particular we introduce classes of CBFs inducing non trivial \textit{damping injection} actions for mechanical systems, able to achieve richer behaviours than mere stabilisation of equilibria. We claim this way to give an incremental contribution in equipping the PBC framework with a tool allowing to constructively embed task-oriented specifications in passive designs, often considered over conservative in their basic formulations.

\subsubsection*{Related work} The class of CBFs that preserve passivity include those introduced by the authors in \cite{Singletary2021Safety-CriticalSystems}, which are associate to the so-called \textit{energy-based safety constraints}. This fact, beyond providing a constructive way to guarantee passivity when computing kinematic tasks, reinforces the link between safety-critical and energy-based techniques, a duality stressed in \cite{Ames2019ControlApplications} and explored further in this letter.
Finally we recognise the paper \cite{Capelli2022PassivityEnergy} combining PBC and CBFs, but it is quite unrelated to this work both in technique and motivation: in \cite{Capelli2022PassivityEnergy} safety-critical control is used to passivize a non passive control action taking advantage of the \textit{energy tank} framework \cite{Califano2022OnSystems}. In this letter instead we start with a passive design as nominal controller and study conditions under which safety-critical control preserves passivity. 

In Sec. \ref{sec:background} the background and motivation related to PBC and CBFs are introduced. Sec. \ref{sec:main} presents the result involving passivity preserving safety-critical control, which is specialised to port-Hamiltonian systems in Sec. \ref{sec:pH}. Simulations are presented in Sec. \ref{sec:sim} and Sec. \ref{sec:conc} concludes the paper.

\section{Background}
\label{sec:background}
Consider the affine nonlinear control system:
\begin{equation}
\label{eq:nonlinearaffinesystem}
    \dot{x}=f(x)+g(x)u
\end{equation}
where $x\in \mathcal{D} \subseteq \mathbb{R}^n$ is the state, $u\in \mathcal{U} \subset \mathbb{R}^m$ is the input, $f: \mathbb{R}^{n} \to \mathbb{R}^{n}$ and $g: \mathbb{R}^{n} \to \mathbb{R}^{n \times m}$ are continuously differentiable maps. As a consequence a Lipschitz continuous controller guarantees existence and uniqueness of solutions of (\ref{eq:nonlinearaffinesystem}). 
In the following we briefly introduce the relevant information involving passivity and safety-critical control. We refer to \cite{vanderSchaftL2} for passivity and to \cite{Ames2019ControlApplications,Xu2015RobustnessControl,Ames2017ControlSystems,Singletary2021Safety-CriticalSystems} for safety-critical control for references which completely cover the presented background.

\subsection{Passivity and passivity-based control}

\subsubsection*{Passivity}

A system in the form (\ref{eq:nonlinearaffinesystem}) equipped with an output $y\in \mathcal{Y} \subset \mathbb{R}^m$, is said to be \textit{passive} with respect to a differentiable \textit{storage function} $S : \mathcal{D} \to \mathbb{R}^+$ and input-output pair $(u,y)$, if the following inequality holds $\forall u \in \mathcal{U}$:

\begin{equation}
\label{eq:passivity}
 \dot{S}=L_{f}S(x)+L_gS(x)u\leq y^{\top}u,
\end{equation}
where $L_{f}S(x):=\frac{\partial S}{\partial x} ^{\top}f(x)\in \mathbb{R}$, $L_gS(x):=\frac{\partial S}{\partial x} ^{\top}g(x)\in \mathbb{R}^{1 \times m}$ and the gradient of $S(x)$ is $\frac{\partial S}{\partial x} \in \mathbb{R}^{n}$.

For physical systems, where $S(x)$ represents energy and $y^{\top} u$ power flow, condition (\ref{eq:passivity}) is a statement of energy conservation, i.e., the variation of energy in the system is bounded by the power flowing in the system. The inequality margin in (\ref{eq:passivity}) is due to the natural \textit{dissipation} of the system $d(x)=-L_{f}S(x)$. An equivalent condition for (\ref{eq:passivity}) of system (\ref{eq:nonlinearaffinesystem}) with output $y$ is then $d(x)\geq 0$ and $y=L_g S(x)^{\top}$.

\subsubsection*{Passivity-based control}
\begin{figure}[h]
  \hspace{-10mm}
    \includegraphics[scale=0.60]{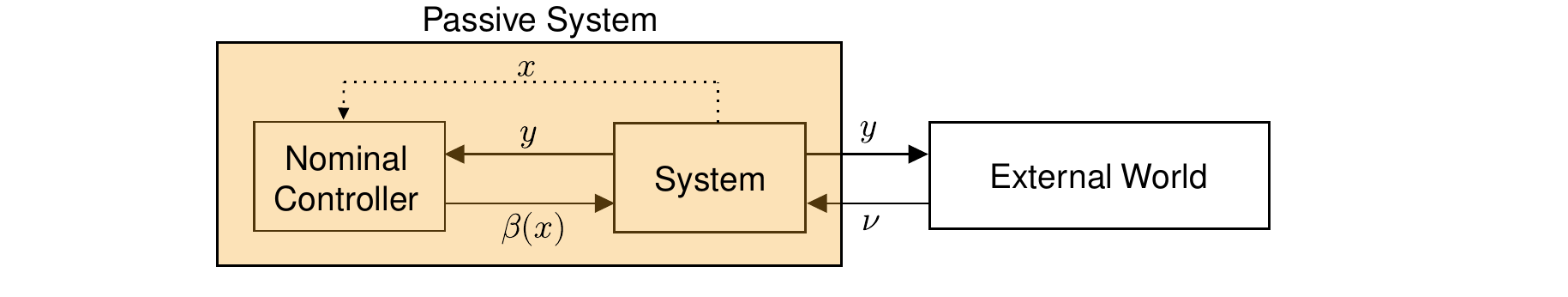}
    \caption{The interconnection view of passivity}
    \label{fig:passivity}
\end{figure}

Passivity-based control (PBC) aims to design a controller for a system in the form (\ref{eq:nonlinearaffinesystem}) in such a way that the closed-loop system is passive. We refer to \cite{Duindam2009ModelingSystems,Ortega2004InterconnectionSurvey,vanderSchaftL2,Stramigioli2015Energy-AwareRobotics,Califano2022OnSystems} for in depth motivations underlying passive designs, but in brief we recognise two distinct motivations. \textit{New methods to design stabilising controllers:} Stability is a corollary of passivity under weak conditions qualifying storage functions as Lyapunov functions. The framework of PBC proposes new methodologies to constructively build those functions with arguments involving the performance of desired closed-loop systems, and not only stability \cite{Ortega2002InterconnectionSystems,Ortega2001PuttingControl};
\textit{Robust stability:} Passive controllers represent a feasible solution to make the closed-loop system robustly stable to unknown environmental interactions, i.e., passive designs are necessary for stability when the controlled system interacts with other unknown passive systems \cite{Stramigioli2015Energy-AwareRobotics,Capelli2022PassivityEnergy,benzi22}. This concept is depicted in Fig. \ref{fig:passivity}: if a complete level of agnosticity with respect to the ``external world" system is required (because e.g., the controlled system evolves in unstructured scenario), then a passive closed-loop system guarantees that when it interfaces with a physical (passive) system, the interconnection is stable.


We consider the case of a passive controller realised by means of a state feedback law. In particular the PBC objective for system (\ref{eq:nonlinearaffinesystem}) is to find a state feedback law $u(x)=\beta(x)+\nu$ such that the closed-loop system 

\begin{equation}
\label{eq:passivecl}
    \begin{cases}
      \dot{x}= f_{cl}(x)+g(x)\nu\\
      y=g(x)^{\top}\frac{\partial S_{cl}}{\partial x}
    \end{cases} 
\end{equation}
is passive with respect to a closed loop storage function $S_{cl}(x)$ and input-output pair $(\nu,y)$, where $f_{cl}(x)=f(x)+g(x)\beta(x)$, see Fig. \ref{fig:passivity}. Notice that in this case passivity reduces to $0 \leq d_p(x):=-L_{f_{cl}}S_{cl}(x)$, i.e., the natural dissipation of the passively controlled system has to be non negative. 
The performance of the controlled system along a task are encoded in the particular way $S_{cl}(x)$ and $f_{cl}(x)$ are allowed to be chosen, which in general requires solving matching PDEs \cite{Ortega2002InterconnectionSystems,Duindam2009ModelingSystems}. However some significant particular cases which can be conveniently addressed by means of these design methods encompass e.g., all potential compensation techniques for mechanical systems (falling in the so-called \textit{energy balance} (EB-PBC) methods), which will be treated in the sequel as a case study. 

\subsection{Control-barrier functions and safety-critical control}

Control barrier functions represent a technique to guarantee forward invariance of a set $\mathcal{C}$, normally called \textit{safe set}, i.e., the control goal is to design a state feedback $u(x)=k(x)$ for system (\ref{eq:nonlinearaffinesystem}) resulting in the closed-loop system $\dot{x}=f_{cl}(x)=f(x)+g(x)k(x)$ such that

\begin{equation}
    \forall x(0) \in \mathcal{C} \implies x(t)\in \mathcal{C} \,\,\, \forall t>0.
\end{equation}
The safe set $\mathcal{C}$ is built as the superlevel set of a continuously differentiable function $h:\mathcal{D}\to \mathbb{R}$, i.e.,

\begin{align*}
    \mathcal{C} = \{ x\in \mathcal{D} : h(x)\geq0 \}, \\
     \partial \mathcal{C} = \{ x\in \mathcal{D} : h(x)=0 \}, \\
      \textrm{Int} (\mathcal{C}) = \{ x\in \mathcal{D} : h(x)>0 \}.
\end{align*}
The function $h(x)$ is then a \textit{control barrier function} (CBF) on $\mathcal{D}$ if $\frac{\partial h}{\partial x}(x)\neq 0, \forall x \in \partial \mathcal{C}$ and 

\begin{equation}
 \sup_{u \in \mathcal{U}} \dot{h}(x,u)=\sup_{u \in \mathcal{U}} [ L_{f}h(x)+L_g h(x)u ] \geq -\alpha(h(x))   
\end{equation}
for all $x\in \mathcal{D}$ and some \textit{extended class $\mathcal{K}$ function\footnote{A function $\alpha: (-b,a) \to (- \infty, \infty)$ with $a,b>0$, which is continuous, strictly increasing, and $\alpha(0)=0$.}} $\alpha$. 
The following key result connects the existence of such CBF to forward invariance of the corresponding safe set.

\begin{theorem}[\cite{Ames2017ControlSystems}]
Let $h(x)$ be a CBF on $\mathcal{D}$ for (\ref{eq:nonlinearaffinesystem}). Any locally Lipschitz controller $u(x)=k(x)$ such that $L_{f}h(x)+L_g h(x)k(x) \geq - \alpha(h(x))$ provides forward invariance of the safe set $\mathcal{C}$. Additionally the set $\mathcal{C}$ is asymptotically stable on $\mathcal{D}$.
\end{theorem}
The way controller synthesis induced by CBFs are implemented is to use them as \textit{safety filters}, transforming a desired state-feedback control input $u_{\textrm{des}}(x)$ into a new state-feedback control input $u^*(x)$ in a minimally invasive fashion in order to guarantee forward invariance of $\mathcal{C}$. In practice, the following Quadratic Program (QP) is solved:

\begin{equation}
\label{eq:LQ}
\begin{aligned}
u^*(x)=\argmin_{u\in \mathcal{U}} \quad & ||u-u_{\textrm{des}(x)} ||^2\\
\textrm{s.t.} \quad & L_{f}h(x)+L_g h(x)u \geq - \alpha(h(x))
\end{aligned}
\end{equation}

The transformation of the desired control input $u_{\textrm{des}}(x)$ in $u^*(x)$ by solving (\ref{eq:LQ}) is denoted as \textit{safety-critical control}, or \textit{safety-critical filtering}.
A last result that will be crucially used in this work is the following lemma.
\begin{lemma}[\cite{Xu2015RobustnessControl,Singletary2021Safety-CriticalSystems}]
Let $h(x)$ be a CBF on $\mathcal{D}$ for (\ref{eq:nonlinearaffinesystem}) and assume $\mathcal{U}=\mathbb{R}^{m}$ and $L_{g}h(x)\neq 0, \,\, \forall x \in \mathcal{D}$. Define $\Psi(x;u_{\textup{des}})=\dot{h}(x,u_{\textup{des}}(x))+\alpha(h(x))$. A closed-form solution for (\ref{eq:LQ}) is given by $u^*(x)=u_{\textup{des}}(x)+u_{\textup{safe}}(x)$, where
\begin{equation}
\label{eq:safetyComponent}
u_{\textup{safe}}(x)= 
    \begin{cases}
    \begin{aligned}
      &-\frac{L_g h(x)^{\top}}{L_g h(x) L_g h(x)^{\top}}\Psi(x;u_{\textup{des}}) &\textup{if} \, \Psi(x;u_{\textup{des}})<0 \\
      &0 &\textup{if} \,  \Psi(x;u_{\textup{des}})\geq 0   
      \end{aligned}
      \end{cases} 
\end{equation}
\end{lemma}

\begin{Note}[\textbf{Disclaimer on the term ``safety''}]
In the following we refer to ``safety" for CBF-related terminology (e.g., safety-critical filtering, safe set, etc.). We stress that this concept of safety is in general not connected to safety guarantees in the sense of preventing physical safety hazards (e.g., human-robot collisions), which are often characterised by fixed thresholds in the amount of admissible energy or power transfer \cite{Tadele2014CombiningRobots}. CBF-related designs can nevertheless be very useful to deal with this latter type of safety, which we will refer to as ``physical safety" in the sequel.
\end{Note}

\section{Passivity Preserving Safety-Critical Control}
\label{sec:main}
\begin{figure}[h]
  \hspace{-5mm}
    \includegraphics[scale=0.60]{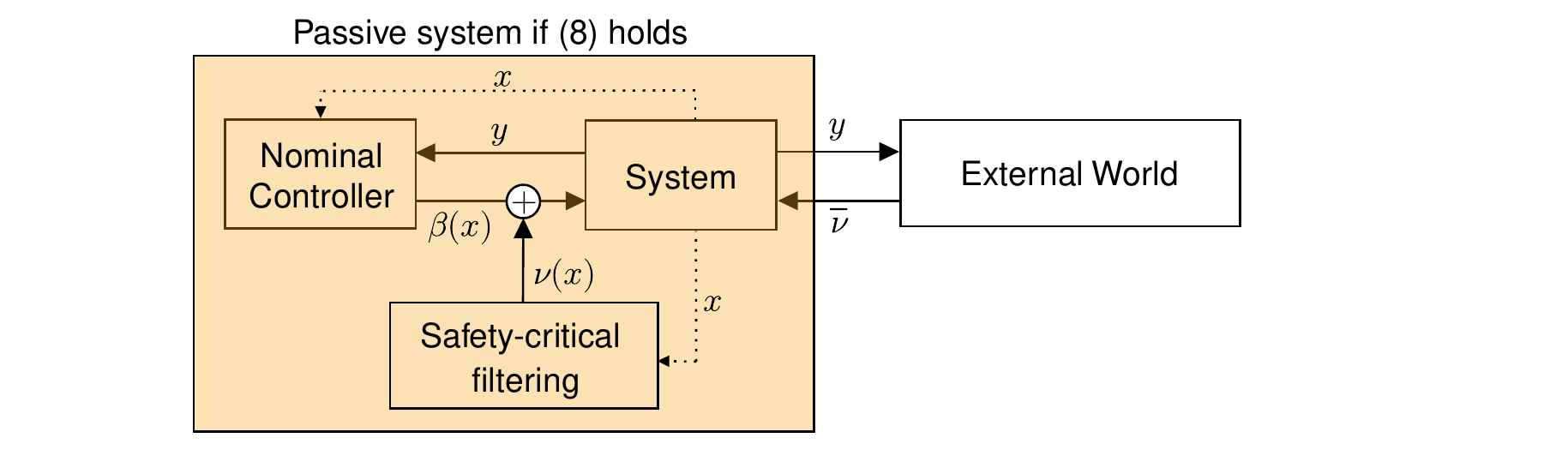}
    \caption{Graphical support to Theorem \ref{th:main}.}
    \label{fig:passivity1}
\end{figure}

In this section we investigate under which conditions passivity of (\ref{eq:passivecl}) is preserved under safety-critical filtering. This will characterise a class of CBFs, which might be useful for different reasons (e.g., physical safety, obstacle avoidance, etc.), that can be used to filter \textit{a posteriori} a passive action without compromising passivity of the new closed-loop system. 
The following theorem, graphically supported by Fig. \ref{fig:passivity1}, provides the result. 
\begin{theorem}
\label{th:main}
Let system (\ref{eq:nonlinearaffinesystem})  with $u(x)=\beta(x)+\nu$ result in the passive closed-loop system (\ref{eq:passivecl}). A safety-filtering on (\ref{eq:passivecl}) induced by a CBF $h(x)$, results in the new controller $u(x)=\beta(x)+\nu(x)+\bar{\nu}$. We indicate with $d_p(x)=-L_{(f+g\beta)} S_{cl}(x)$ the dissipation of the passive system (\ref{eq:passivecl}) and $\Psi(x;\beta)=\dot{h}(x,\beta(x))+\alpha(h(x))$. The resulting closed-loop system is passive with respect to $S_{cl}(x)$ and $(\bar{\nu},y)$ if and only if 

\begin{equation}
\label{eq:condition}
    -\frac{L_g S_{cl}(x) L_g h(x)^{\top} } {L_g h(x) L_g h(x)^{\top} } \Psi(x;\beta) \leq d_p(x)
\end{equation}
when $\Psi(x;\beta)<0$. Furthermore, independently whether passivity is preserved, the instantaneous power that the safety-critical controller injects in the system is given by the left hand side of (\ref{eq:condition}) when $\Psi(x;\beta)<0$.

\end{theorem}
\begin{proof}
The task is to check when the system 
\begin{equation}
\label{eq:passiveclcl}
    \begin{cases}
      \dot{x}= f(x)+g(x)\beta(x) +g(x)\nu(x)+g(x)\bar{\nu}\\
      y=g(x)^{\top}\frac{\partial S_{cl}}{\partial x}
    \end{cases} 
\end{equation}
is passive with respect to $S_{cl}(x)$ and the input-output pair $(y,\bar{\nu})$, where the desired input in (\ref{eq:LQ}) and (\ref{eq:safetyComponent}) is $u_{\text{des}}(x)=\beta(x)$ and the safety component in (\ref{eq:safetyComponent}) is $u_{\textrm{safe}}(x)=\nu(x)$. Due to the available closed-form solution (\ref{eq:safetyComponent}) we can directly calculate the dissipation inequality for (\ref{eq:passiveclcl}):
\begin{align*}
    \dot{S}_{cl}=-d_p(x)+L_g S_{cl}(x)\nu(x)+L_{g}S_{cl}(x)\bar{\nu}.
\end{align*}
Passivity condition $\dot{S}_{cl}\leq y^{\top}\bar{\nu}$ holds if and only if $L_g S_{cl}(x)\nu(x)\leq d_p(x)$, where $L_g S_{cl}(x)\nu(x)$ is the power the safety-critical controller injects in the system. The case $\Psi(x;\beta)\geq 0$ is always satisfied since $\nu(x)=0$ and $d_p(x)\geq 0$ because of passivity of (\ref{eq:passivecl}), while the case $\Psi(x;\beta)< 0$ corresponds to (\ref{eq:condition}), which concludes the proof.
\end{proof}

\section{EB-PBC for port-Hamiltonian systems with Safety-Filtering}
\label{sec:pH}
In this section we specialise the result to mechanical systems and without loss of generality we use a port-Hamiltonian (pH) formulation to describe their dynamics \cite{Duindam2009ModelingSystems}. This modeling technique is often used in the development of PBC schemes since it explicitly encodes the energetic structure of the underlying physical systems. One of the contributions of this section is to use this formulation in the CBF framework. We will show how several manipulations, especially involving the so-called \textit{energy-based safety constraints} \cite{Singletary2021Safety-CriticalSystems} (and their generalisation introduced in the sequel), will gain intuitive and technical advantage.

\subsection{Port-Hamiltonian systems and EB-PBC}

The input--state--output representation of a port--Hamiltonian system consists in an instance of (\ref{eq:nonlinearaffinesystem}) with output $y\in \mathbb{R}^{m}$ in the form: 
\begin{equation}
\label{eq:pHopenloop}
    \begin{cases}
      \dot{x}= (J(x)-R(x))\frac{\partial H}{\partial x}+g(x)u \\
      y=g(x)^{\top}\frac{\partial H}{\partial x}
    \end{cases} 
\end{equation}
where $J(x)=-J(x)^{\top}$ and $R(x)=R(x)^{\top}\geq 0$ are respectively skew-symmetric and positive semi-definite symmetric matrices representing the power-preserving and the dissipative components of the system. The non-negative function $H: \mathcal{D}\to \mathbb{R}^{+}$ is called the \textit{Hamiltonian} and maps the state into the total physical energy of the system. As a matter of fact system (\ref{eq:pHopenloop}) is passive by construction with storage function $H(x)$ and input-output pair $(y,u)$ since, using skew-symmetry of $J(x)$, positive-definitness of $R(x)$ and indicating with $f(x)=(J(x)-R(x))\frac{\partial H}{\partial x}$:

\begin{equation}
\label{eq:passivitypH}
    \dot{H}=L_f H(x)+y^{\top}u=-\frac{\partial H}{\partial x}^{\top}R(x)\frac{\partial H}{\partial x}+y^{\top}u\leq y^{\top}u
\end{equation}
which is a statement of energy conservation.
PBC techniques in this framework are conveniently addressed by designing a target closed-loop system in port-Hamiltonian form and ``matching" it to the open-loop port-Hamiltonian system with a parametrised feedback law $u(x)=\beta(x)+\nu$. A complete description of these design methodology can be found in \cite{Ortega2002InterconnectionSystems,Ortega2008ControlSystems}. In the following we address a particular case, which encompasses many control schemes of interest, referred to as \textit{energy balancing} (EB)-PBC.

\begin{theorem}(\cite{Ortega2008ControlSystems}) 
\label{th:EBPBC}Consider the open-loop system (\ref{eq:pHopenloop}) undergoing its energy balance (\ref{eq:passivitypH}) where we indicate with $d(x)=\frac{\partial H}{\partial x}^{\top}R(x)\frac{\partial H}{\partial x}$ the natural dissipation. 
If it is possible to find $\beta(x)$ such that $\dot{\Bar{V}}(x) = y^{\top}\beta(x)$ where $\Bar{V}(x)=S_{cl}(x)-H(x)$, then the control law $u(x) = \beta(x)+\nu$ is such that  $\dot{S}_{cl}(x) = y^{\top}\nu -d(x)$ is satisfied, i.e., a passive closed-loop system with storage function $S_{cl}(x)$ and input output port $(y,\nu)$ is obtained. The closed-form solution of the EB-PBC controller is given by 
\begin{equation}
\label{eq:closedformEB}
    \beta(x) = g^+(x)(J(x)+R(x))\frac{\partial \Bar{V}}{\partial x}
\end{equation}
where $g^+(x)$ is the left pseudo--inverse of $g(x)$ and $\bar{V}(x)$ satisfies the following matching equations 
\begin{equation}\label{eq:matching_cond}
    \begin{bmatrix}
        g^{\perp}(x)(-J(x)+R(x))\\
        g^\top(x)
    \end{bmatrix}
    \frac{\partial \Bar{V}}{\partial x} = 0
\end{equation}
being $g^{\perp}(x)$ is a left full--rank annihilator of $g(x)$.
\end{theorem}
The design procedure is complemented with some desired properties on the closed-loop storage function $S_{cl}(x)$, normally by choosing the minimum of $S_{cl}(x)$ as the point that needs to be stabilised. The EB-PBC design is then possibly completed by the so-called \textit{damping injection} procedure, where a negative output feedback on $\nu=-D_i y$ for some positive definite matrix $D_i$ increases the convergence rate to the minimum of $S_{cl}(x)$. 

The class of obtained passive closed-loop systems can directly undergo a the safety filtering through a CBF $h(x)$ on which condition (\ref{eq:condition}) determines whether passivity is preserved. 

\subsection{Mechanical systems}
In order to better comprehend the proposed methodology and discuss how it complements standard EB-PBC approaches, let us specialise system (\ref{eq:pHopenloop}) to mechanical systems, where we introduce the state $x=(q^{\top},p^{\top}) \in \mathbb{R}^{2n}$ as canonical Hamiltonian coordinates on the cotangent bundle of the configuration manifold of the system. 
Let $q\in \mathbb{R}^{n}$ be the vector of generalized coordinates. $p\in \mathbb{R}^n$ denotes the generalized momenta, $p:=M(q)\dot q$, where $M(q)=M(q)^\top>0$ is the positive definite inertia matrix of the system. The equations of motions in canonical form are given by \eqref{eq:pHopenloop} with 
$$
    J(x)-R(x) = \begin{bmatrix}
            0 & I_n\\
            -I_n & -D
        \end{bmatrix}, ~g(x) =\begin{bmatrix}
            0\\
            B
        \end{bmatrix}
$$
resulting in 
\begin{equation}\label{eq:mech_ph}
\begin{cases}
    \begin{aligned}
        \begin{bmatrix}
            \dot q\\
            \dot p
        \end{bmatrix}&=
        \begin{bmatrix}
            0 & I_n\\
            -I_n & D
        \end{bmatrix}
        \begin{bmatrix}
            \frac{\partial H}{\partial q}\\
            \frac{\partial H}{\partial p}
        \end{bmatrix} +
        \begin{bmatrix}
            0\\
            B
\        \end{bmatrix}u\\
        y &= \begin{bmatrix}
            0 & B^{\top}
        \end{bmatrix} \begin{bmatrix}
            \frac{\partial H}{\partial q}\\
            \frac{\partial H}{\partial p}
        \end{bmatrix} =B^{\top} \dot{q}
    \end{aligned}
    \end{cases}
\end{equation}
where $H:\mathbb{R}^{2n} \to \mathbb{R}$ is the total energy (Hamiltonian)
\[
    H(q,p)= \frac{1}{2}p^\top M^{-1}(q)p + V(q),
\]
$V:\mathbb{R}^{n} \to \mathbb{R}$ maps the position state to conservative potentials (gravity, elastic effects), $D=D^\top \geq 0$ takes into account dissipative and friction effects, $B\in \mathbb{R}^{n \times n}$ is the input matrix\footnote{For lightening notation we hide possible state dependencies on $D$ and $B$.}, $I_n$ indicates the $n \times n$ identity matrix and non specified dimensions of matrices, comprising those with only zero entries and denoted with the symbol ``$0$", are clear from the context. 

The EB-PBC procedure applied to (\ref{eq:mech_ph}) encompasses all passive potential compensation techniques for mechanical systems. In fact, with a look at Theorem \ref{th:EBPBC}, and considering for simplicity $\textrm{rank}(B)=n$, it is easy to see that if one lets the so called \textit{added energy} $\bar{V}$ to depend only on the position state $q$, then the matching conditions (\ref{eq:matching_cond}) are automatically satisfied and the control (\ref{eq:closedformEB}) reduces to $\beta(q)=-\frac{\partial \bar{V}}{\partial q}$.
This procedure, which will be considered from now on, can be used to \textit{de facto} re-derive PD+potential compensation controllers with novel arguments (see \cite{massaroli2021optimal}), by choosing $\bar{V}(q)=-V(q)+\frac{1}{2}q^{\top}Kq$ with $K=K^{\top}\geq 0$, and add damping injection to increase the convergence to the minima of the closed-loop storage function 
\begin{equation}
    S_{cl}(q,p)=H(q,p)+\bar{V}(q).
\end{equation}
More generally, any choice of $\bar{V}(q)$ which is bounded from below\footnote{Boundedness of $\bar{V}(q)$ qualifies $S_{cl}(q,p)$ as a valid storage function.} gives raise to a passive closed-loop system, 
as an instance of (\ref{eq:passivecl}) in the form
\begin{equation}\label{eq:mech_ph_passive}
\begin{cases}
    \begin{aligned}
        \begin{bmatrix}
            \dot q\\
            \dot p
        \end{bmatrix}&=
        \begin{bmatrix}
            0 & I_n\\
            -I_n & D
        \end{bmatrix}
        \begin{bmatrix}
            \frac{\partial S_{cl}}{\partial q}\\
            \frac{\partial S_{cl}}{\partial p}
        \end{bmatrix} +
        \begin{bmatrix}
            0\\
            B
\        \end{bmatrix}\nu\\
        y &= \begin{bmatrix}
            0 & B^{\top}
        \end{bmatrix} \begin{bmatrix}
            \frac{\partial S_{cl}}{\partial q}\\
            \frac{\partial S_{cl}}{\partial p}
        \end{bmatrix} =B^{\top} \dot{q}.
    \end{aligned}
    \end{cases}
\end{equation}
With a slight abuse of notation we denote $D$ in (\ref{eq:mech_ph_passive}) the dissipation matrix that possibly includes a damping injection component, and consistently with the notation in Theorem \ref{th:main} we indicate $\bar{d}_p(q,p)=\frac{\partial h}{\partial p}^{\top}D\frac{\partial S_{cl}}{\partial p}$.
We now apply the results of Theorem \ref{th:main} to system (\ref{eq:mech_ph_passive}), giving an energetic interpretation of safety-critical filtering on passively controlled mechanical systems. We indicate with $P_{\textrm{safe}}(x)=L_g S_{cl}(x)\nu(x)$, the power injected by the safety filtering component of the controller, and with $\{\cdot,\cdot \}$ the \textit{Poisson bracket} induced by the symplectic structure canonically present in mechanical systems, i.e., for two smooth real-valued functions $\phi(q,p),\xi(q,p)$, the Poisson bracket is defined as 
$\{ \phi, \xi \}=\frac{\partial \phi}{\partial q} \frac{\partial \xi}{\partial p}-\frac{\partial \phi}{\partial p} \frac{\partial \xi}{\partial q}$. We use the notation $P_{\textrm{safe}}|_{\Psi < 0}$ to indicate the power injected by the safety-critical controller when $\Psi < 0$, since otherwise $P_{\textrm{safe}}=0$.

Applying safety-critical filtering induced by a CBF $h(q,p)$ to (\ref{eq:mech_ph_passive}), one obtains:


\begin{equation}
    \Psi(q,p;\beta)=\{h,S_{cl}\} + \bar{d}_p(q,p)+ \alpha(h(q,p))
\end{equation}

\begin{equation}
   P_{\textrm{safe}}|_{\Psi < 0}= -\frac{\dot{q}^{\top} B B^{\top} \frac{\partial h}{\partial p}}{\frac{\partial h}{\partial p}^{\top} B B^{\top} \frac{\partial h}{\partial p}} \Psi(q,p;\beta)   
\end{equation}
and we remind that the condition (\ref{eq:condition}) for passivity preservation is $P_{\textrm{safe}}|_{\Psi < 0}\leq d_p(q,p)$.

We now consider the class of candidate CBFs in the form 
\begin{equation}
\label{eq:genEnCBF}
    h(q,p)=-K_e(q,p)+\alpha_E \bar{h}(q)+\bar{E},
\end{equation}
where $K_e(q,p)=\frac{1}{2}p^\top M^{-1}(q)p$ is the kinetic energy, $\bar{h}(q)$ is a smooth function on the position variable only, $\bar{E}\in \mathbb{R}^{+}$ and $\alpha_E \in \mathbb{R}$. We call the superlevel sets of CBFs in the form (\ref{eq:genEnCBF}) \textit{generalised energy-based safe sets}, see Remark \ref{rem:safetykinematics}, and present the following corollary.

\begin{corollary}
\label{cor:cor}
Every candidate CBF in the form (\ref{eq:genEnCBF}) induces a passivity-preserving safety-critical filtering. Furthermore the dissipated power by the controller is always negative when $\Psi(q,p;\beta)<0$ and equals the constraint value, i.e., $P_{\textrm{safe}}|_{\Psi < 0}=\Psi(q,p;\beta)$. Furthermore $\Psi(q,p;\beta)=\{\alpha_{E}\bar{h}+V^t,K_e\}+\bar{d}_p(q,p)+\alpha(h(q,p))$ where $V^t(q)=V(q)+\bar{V}(q)$ is the total closed-loop potential.
\begin{proof}
    $\frac{\partial h}{\partial p}=-\frac{\partial K_e}{\partial p}=-\dot{q}$. As a consequence $P_{\textrm{safe}}|_{\Psi < 0}=\Psi(q,p;\beta)$, i.e., condition (\ref{eq:condition}) is always satisfied. The value of the constraint $\Psi(q,p;\beta)$ is easily calculated using skew-symmetry and bilinearity of the Poisson bracket.
\end{proof}
\end{corollary}
The following remarks address particular cases of interest.
\begin{remark}[\textbf{Safety-Critical Kinematic Control}]
\label{rem:safetykinematics}
Using a Lagrangian formalism, in \cite{Singletary2021Safety-CriticalSystems} the authors define the \textit{energy-based safe sets} as the superlevel set of (\ref{eq:genEnCBF}) with $\alpha_E>0$ and $\bar{E}=0$, and prove that it is a valid CBF on its superlevel set. In \cite{Singletary2021Safety-CriticalSystems} the motivation is to implement safety-critical kinematic control, i.e., to make the superlevel set of $\bar{h}(q)$ forward invariant, which cannot be done trivially since $\bar{h}(q)$ is not a valid CBF because $L_{g}\bar{h}=0$ for mechanical systems. The authors prove that with a sufficiently large $\alpha_E$ the superlevel sets of (\ref{eq:genEnCBF}) approach those of $\bar{h}(q)$, and thus solve successfully the safety-critical kinematic control problem. We conclude that all the safety-critical kinematic control schemes developed in \cite{Singletary2021Safety-CriticalSystems} are passivity-preserving since the used CBFs are particular cases of (\ref{eq:genEnCBF}).
\end{remark}

\begin{remark}[\textbf{Physical Safety}]
    Defining safe sets to limit to a constant value the total energy $S_{cl}(q,p)$ or the kinetic energy $K_e(q,p)$ are particular cases of safe sets encoded in (\ref{eq:genEnCBF}) (resp. with $\alpha_E \bar{h}(q)=-V^t(q)$ and $\alpha_E=0$), and thus can be used to impose physical safety constraint along passive designs, as (even if often misunderstood) passivity does not imply physical safety \cite{Califano2022OnSystems}. Notice that when $h(q,p)=-S_{cl}(q,p)+\Bar{E}$, the dissipated power reduces to $P_{\textrm{safe}}|_{\Psi < 0}=- d_p(q,p)+ \alpha(h(q,p))$ since $\{S_{cl},S_{cl}\}=0$.
\end{remark}
We observe that the described safety-filtering procedure provides less conservative ways to implement damping injection schemes on mechanical systems. In fact Corollary \ref{cor:cor} provides conditions under which the safety critical controller acts as a damper, in a different way than a standard derivative action does: the controller, implementing a nontrivial logic encoded in the safety-critical optimisation, damps energy in regions of the state space that conveniently encode task-oriented information though proper choices of CBFs.

\section{Simulations}
\label{sec:sim}

\begin{figure}[h]
\centering
    \includegraphics[scale=1.18]{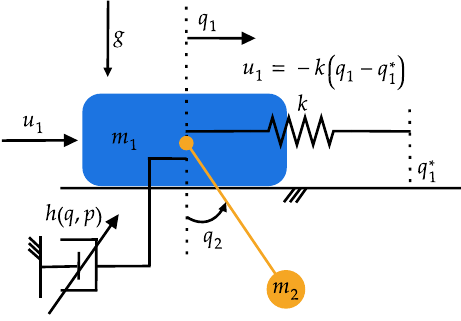}
    \caption{The cart-pole system and the physical representation of its control effects.}
    \label{fig:cart}
\end{figure}
We present simulations involving a cart-pole system, shown in Fig. \ref{fig:cart}. We consider the simple case of a nominal controller implementing a proportional action with reference $q_1^*=1$ on the horizontal coordinate. In the PBC interpretation, the controller acts like a linear spring with stiffness $k$, and the closed-loop system is passive with storage function $S_{cl}(q,p)=H(q,p)+\frac{1}{2}kq_1^2$, being $H(q,p)$ the open-loop Hamiltonian of the system.
To show the role of the passivity-preserving safety-critical controller as a damper, we assume no friction in the plant and no dissipation in the passive controller, i.e., the passively controlled system is \textit{lossless}, a particular case of passivity with $d_p(q,p)=0$. It follows that all the losses in $S_{cl}$ are caused by the safety-critical controller. We perform two classes of simulations with two instances of CBFs in the form (\ref{eq:genEnCBF}), for which Corollary \ref{cor:cor} guarantees that the safety critical controller acts indeed as a damper, as represented in Fig. \ref{fig:cart}. All model parameters are set to unity unless specified, and the initial states of the system are zero both in position and momentum. Furthermore we use $\alpha(h)=\gamma h$ with $\gamma=10 \si{Hz}$\footnote{Notice that the unit of $\gamma$ for CBFs in the form (\ref{eq:genEnCBF}) must be $\si{Hz}$ since the term $\gamma h$ enters as a sum in $\Psi$ which carries the unit of power.}.

\subsubsection{Limiting kinetic energy}
\begin{figure*}[t]
    \centering
    \subfloat{
        \centering
        \includegraphics[width=.71\columnwidth]{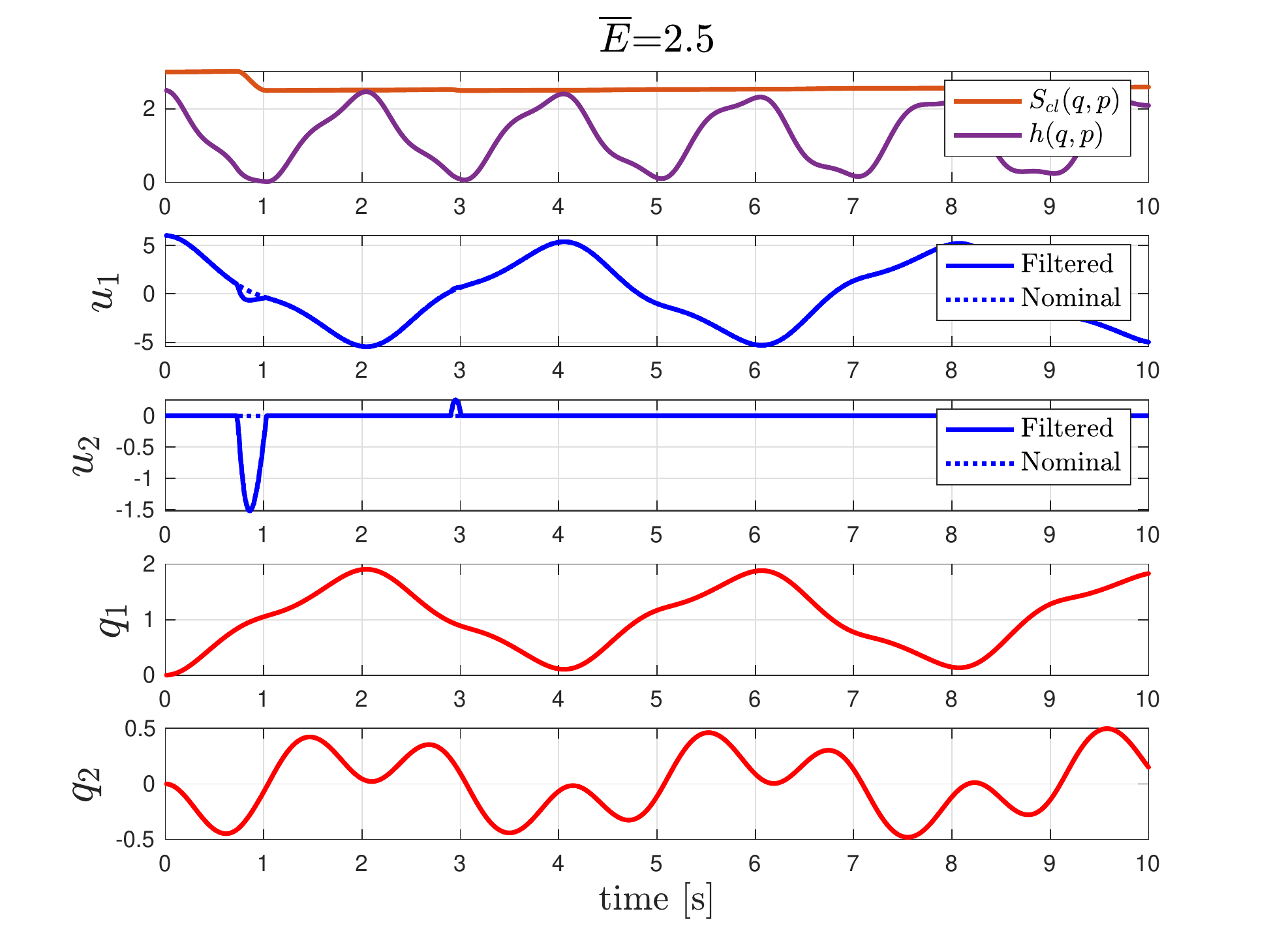}
    } \hspace{-8mm}
    \subfloat{
        \centering
        \includegraphics[width=.71\columnwidth]{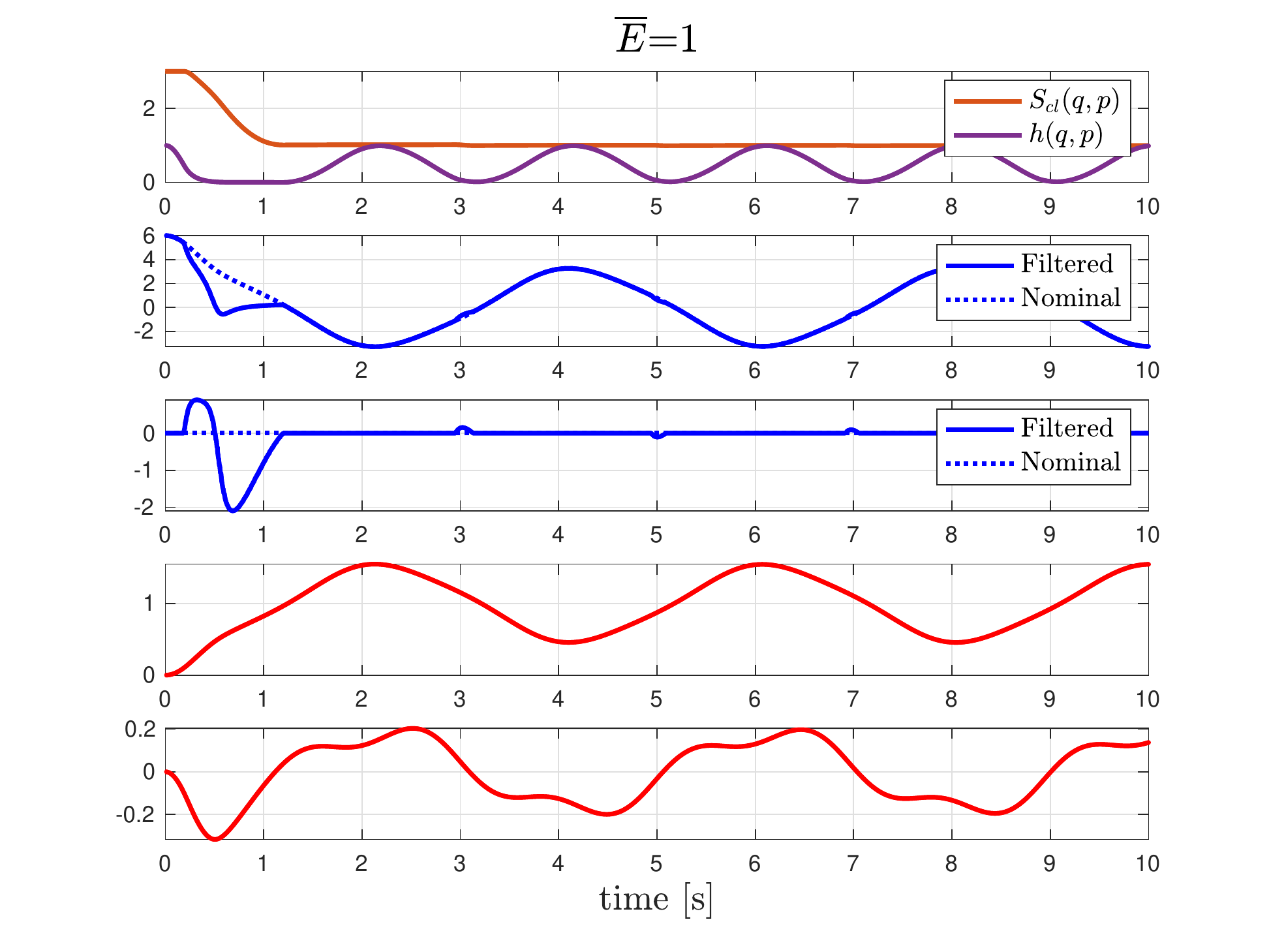}
    }  \hspace{-8mm}
    \subfloat{
        \centering
        \includegraphics[width=.71\columnwidth]{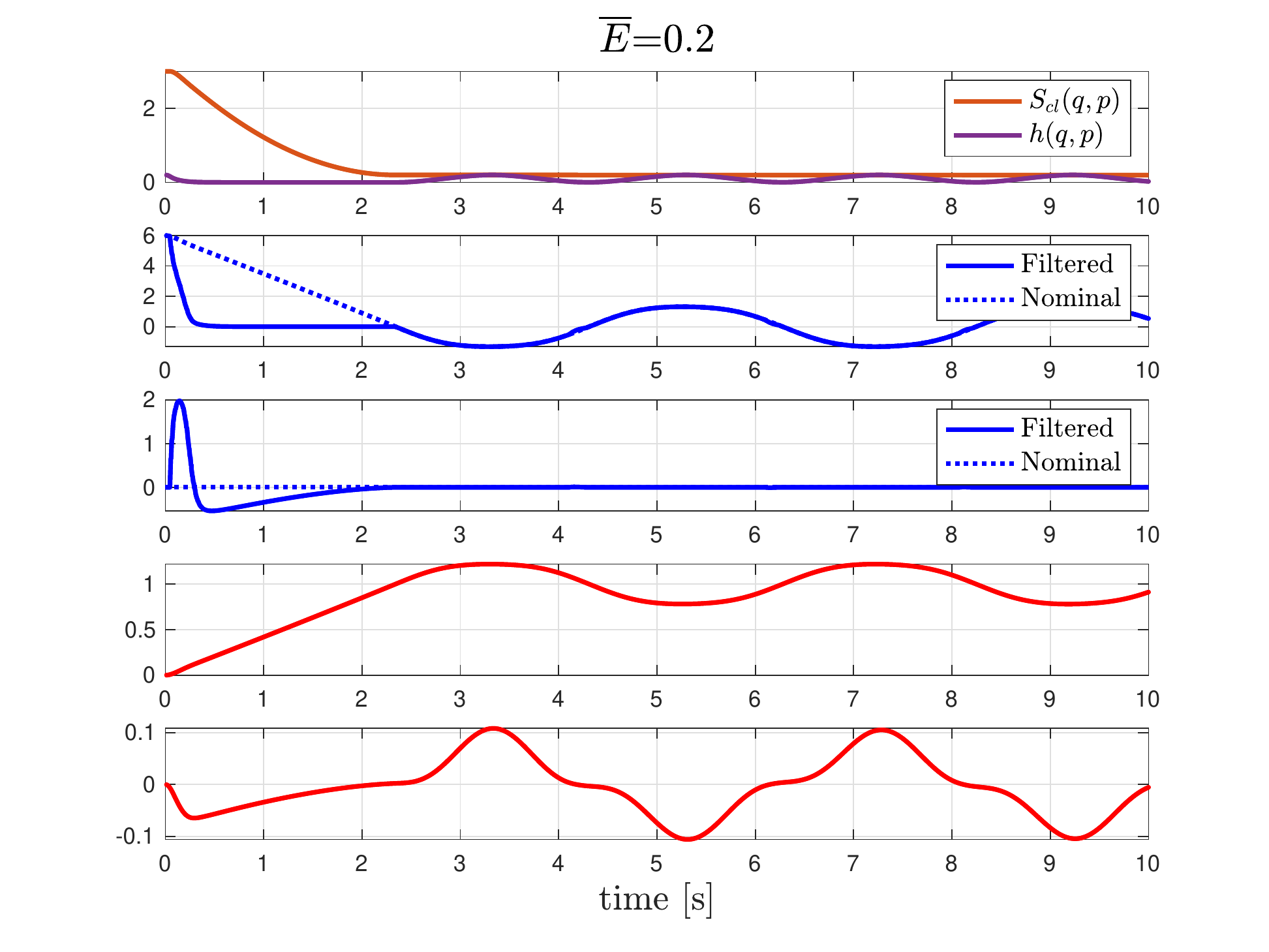}
    }    \caption{Safety critical filtering effect on the lossless system of Fig. \ref{fig:cart} with $h(q,p)=-K_e(q,p)+\bar{E}$ for different $\bar{E}$.}
    \label{fig:kinetikcbf} 
    \vspace{-3mm}
\end{figure*}

Fig. \ref{fig:kinetikcbf} shows the effect of the safety critical controller induced by the CBF $h(q,p)=-K_e(q,p)+\bar{E}$ with different choices of $\bar{E}$, i.e., the safe set is defined in a way to limit the total kinetic energy of the system to a constant value. Since the nominal controller is implemented with $k=6 \si{N/m}$, it results $S_{cl}|_{t=0}=3 \si{J}$, a value that would be nominally conserved along the motion since the system without safety critical filtering is lossless. It is clearly visible that as soon as $h(q,p)$ approaches zero, the safety critical filtering modifies the control action to damp energy from $S_{cl}$. The amplitude of the steady state oscillations around $q_1^*$ decrease when $\bar{E}$ decreases.
\subsubsection{Safety-critical kinematic control}
\begin{figure*}[t]
    \centering
    \subfloat{
        \centering
        \includegraphics[width=.71\columnwidth]{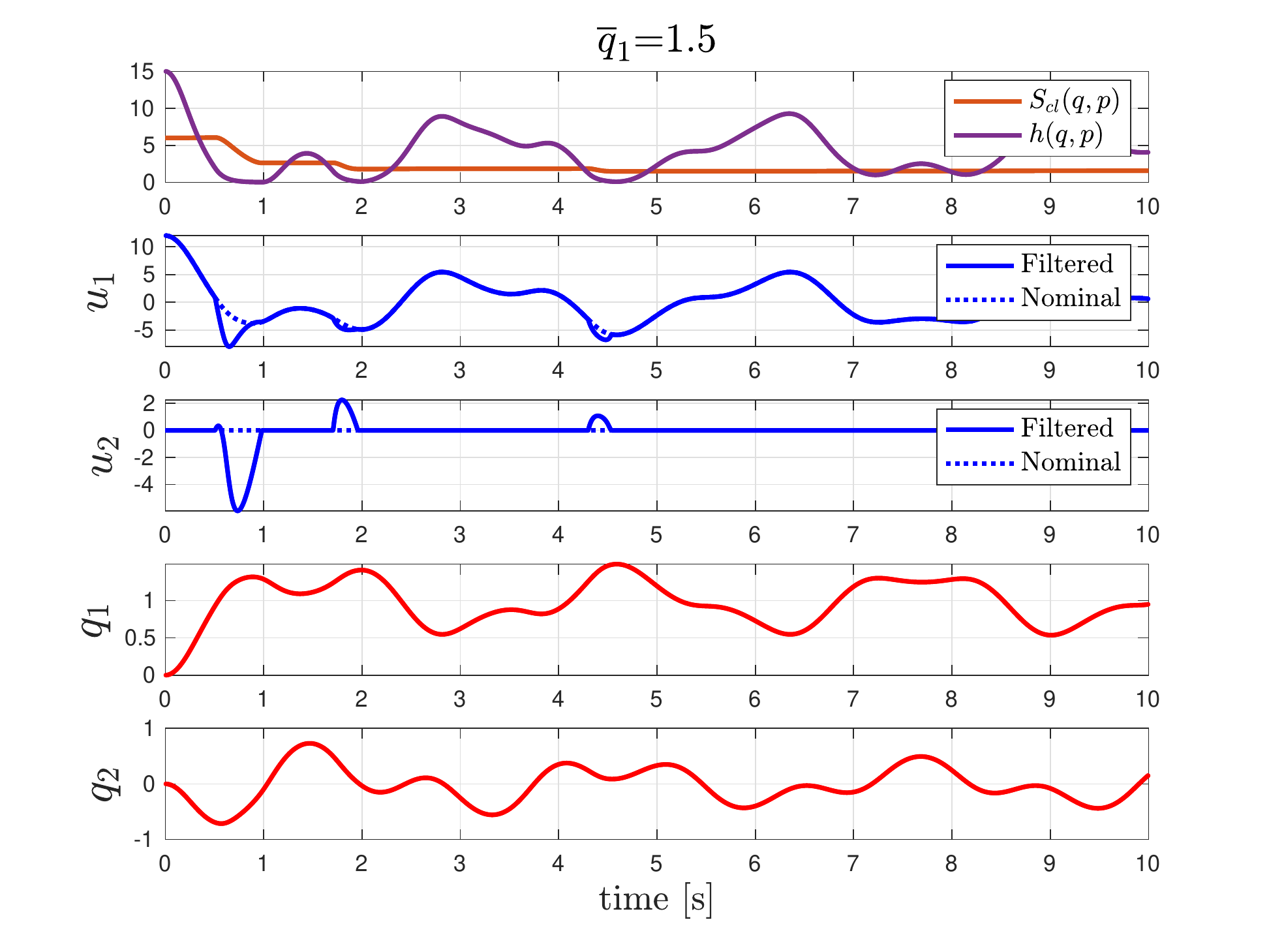}
    } \hspace{-8mm}
    \subfloat{
        \centering
        \includegraphics[width=.71\columnwidth]{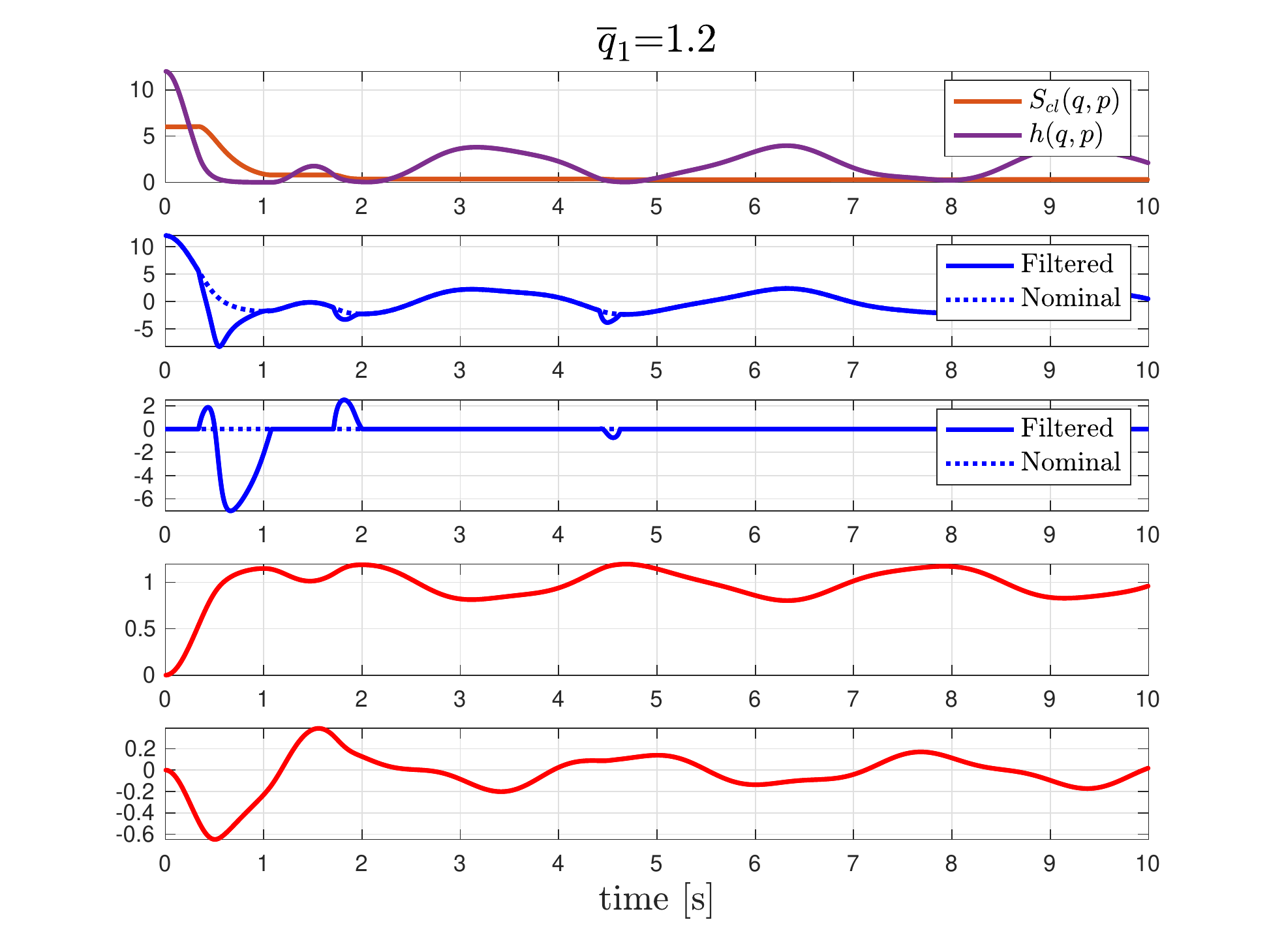}
    }  \hspace{-8mm}
    \subfloat{
        \centering
        \includegraphics[width=.71\columnwidth]{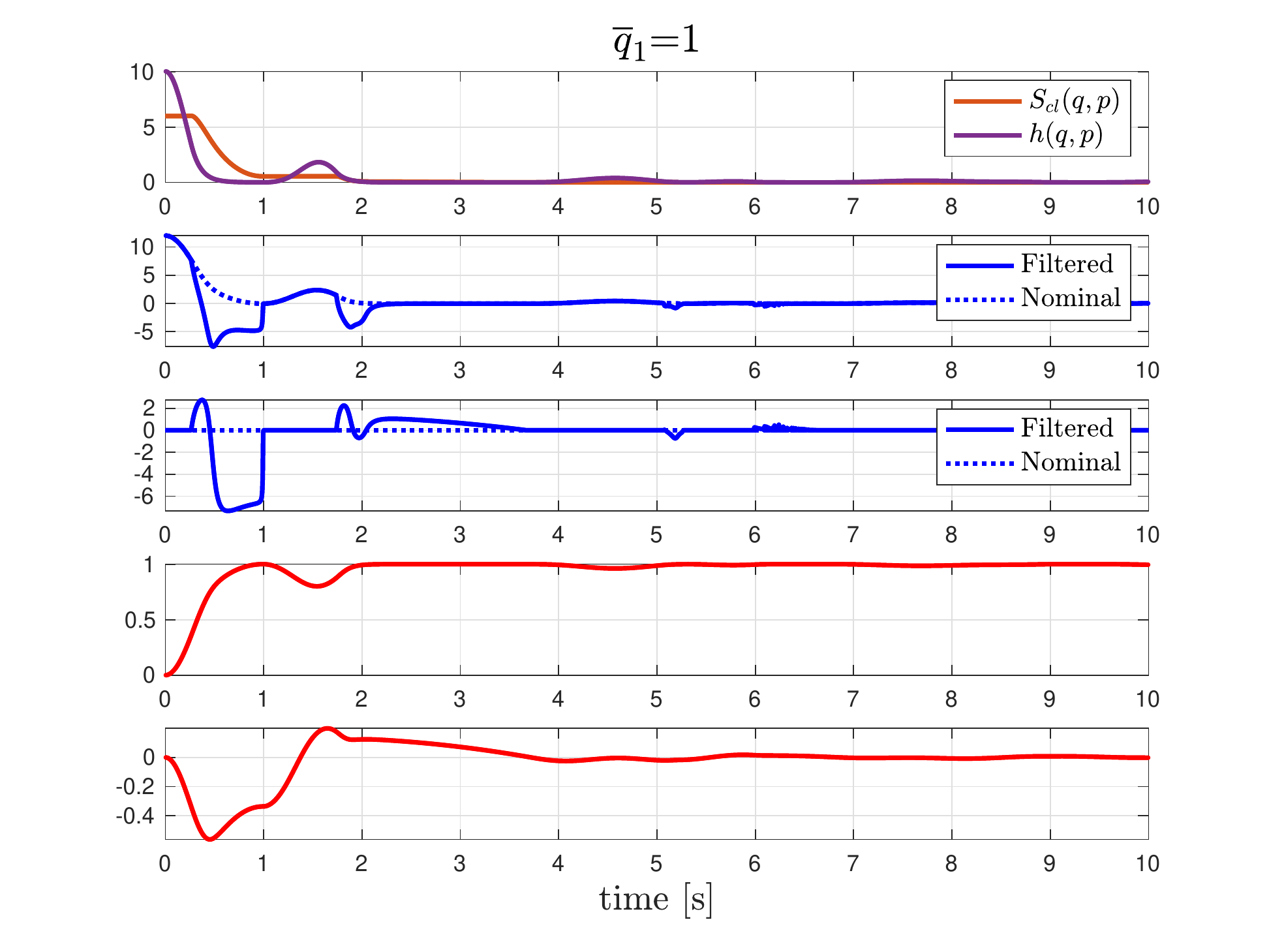}
    }    \caption{Safety critical filtering effect on the lossless system of Fig. \ref{fig:cart} with $h(q,p)=-K_e(q,p)+\alpha_E(\bar{q}_1-q_1)$ for different $\bar{q}$ and $\alpha_E=10$. }
    \label{fig:kinematikcbf} 
    \vspace{-3mm}
\end{figure*}

Fig. \ref{fig:kinematikcbf} shows the results of the simulations with $h(q,p)=-K_e(q,p)+\alpha_E(\bar{q}_1-q_1)$, which approaches (see \cite{Singletary2021Safety-CriticalSystems}) the safe set $q_1\leq \bar{q}_1$ for a sufficiently large $\alpha_E$. As predicted by Corollary \ref{cor:cor}, we observe that the critical safety filtering damps energy from $S_{cl}$ (this time initialised at $6 \si{J}$ since $k=12\si{N/m}$) in a way to constraint the horizontal coordinate to be less than $\bar{q}_1$. 

The experiments prove the concept that it is possible to take advantage of CBFs in the form (\ref{eq:genEnCBF}) to introduce damping effects whose role goes beyond mere stabilisation of equilibria.

\section{Conclusions}
\label{sec:conc}
In this letter we presented conditions under which safety-critical control implemented with CBFs preserves passivity of the underlying system. We specialised the results to mechanical systems in port-Hamiltonian form, which revealed convenient ways to complement passive designs with novel damping injection strategies encoded by generalised energy-based CBFs.

\renewcommand*{\bibfont}{\small}
\printbibliography

\end{document}